\documentclass[12pt]{amsart}

\pdfoutput=1

\usepackage[text={400pt,660pt},centering]{geometry}

\usepackage{esint,amssymb} 
\usepackage{graphicx}
\usepackage{MnSymbol}
\usepackage{mathtools} 
\usepackage[colorlinks=true, pdfstartview=FitV, linkcolor=blue, citecolor=blue, urlcolor=blue,pagebackref=false]{hyperref}
\usepackage{microtype}

\usepackage{bm}
\usepackage{dsfont}

\numberwithin{table}{section}

\begin{document}

\author[J.-C. Mourrat]{J.-C. Mourrat}
\address[J.-C. Mourrat]{DMA, Ecole normale sup\'erieure,
CNRS, PSL University, Paris, France}
\email{mourrat@dma.ens.fr}

\title[On the share of mathematics published by Elsevier and Springer]{On the share of mathematics published \\
by Elsevier and Springer}

\begin{abstract}
For-profit editors such as Elsevier and Springer have been subject to sustained criticism from academics and university libraries, including calls to boycott, and discontinued subscriptions. Mathematicians have played a particularly active role in this critique, and have endeavored to imagine new publication practices and create new journals. This motivates the monitoring of the share of articles published by different editors. I used data from MathSciNet over the period 2000-2017, and focused on the 100 journals with highest citations per article. Within this category, the share of articles published by Elsevier and Springer has steadily increased over this period, from about a third to almost half of the total. 
\end{abstract}

\maketitle

%
%
%
%
%
%

Over the last decades, universities have been confronted with journal price rises that far outpaced the evolution of their budgets \cite{librarians06,dewatripont07,pnas}. 
%
%
%
%
As a consequece, several institutions had to take the difficult decision to interrupt their subscriptions
 \cite{sparc}.
In~2012, more than 10'000 scholars signed a petition declaring their boycott of Elsevier~\cite{cok}. Several alternative publication models have been proposed, where ultimate ownership and control remains in the hands of scholars themselves, and where costs are much lower for authors and readers, or even non-existent \cite{mer,cimpa,epi,fjn,msp}.  

\smallskip

The goal of this note is to try to measure the share of mathematical research papers that get published by large for-profit editors, and how this share has evolved over time. I used data from MathSciNet, an online bibliographic database operated by the American mathematical society, and focused on the period 2000-2017. For each year $n$ and each journal, MathSciNet provides with a number of citations per article, which is computed by counting the number of citations received in year~$n$ by papers published in years $n-5$ to $n-1$, and dividing by the total number of papers published by the journal over this period. For any given year, I focused on the 100 journals with the highest number of citations per article. For each given year~$n$ and each journal in this range, I recorded the number of articles published in years $n-5$ to $n-1$, the number of citations these articles received, and whether, on year~$n$, the journal was edited by Elsevier or Springer, on the one hand, or by another editor, on the other hand.

\begin{figure}
\begin{center}
\includegraphics[scale=0.33]{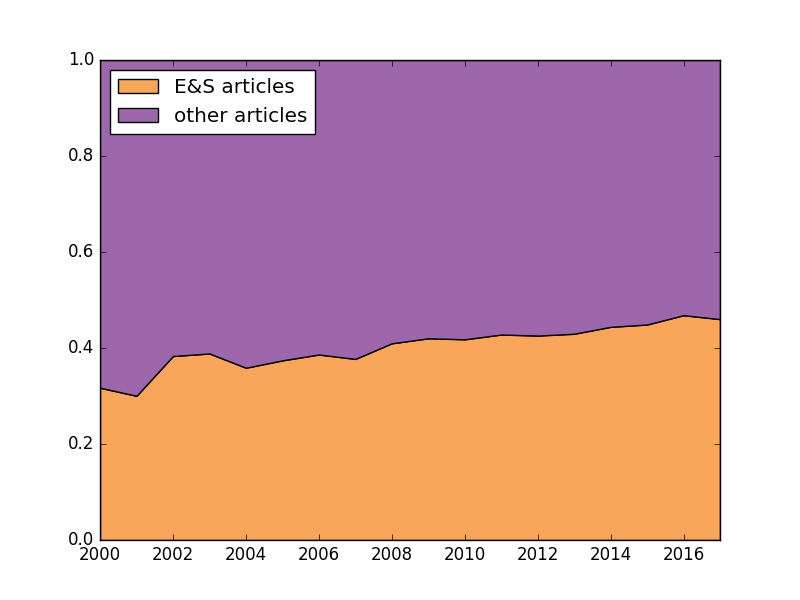}  \hspace{.1cm}
\includegraphics[scale=0.33]{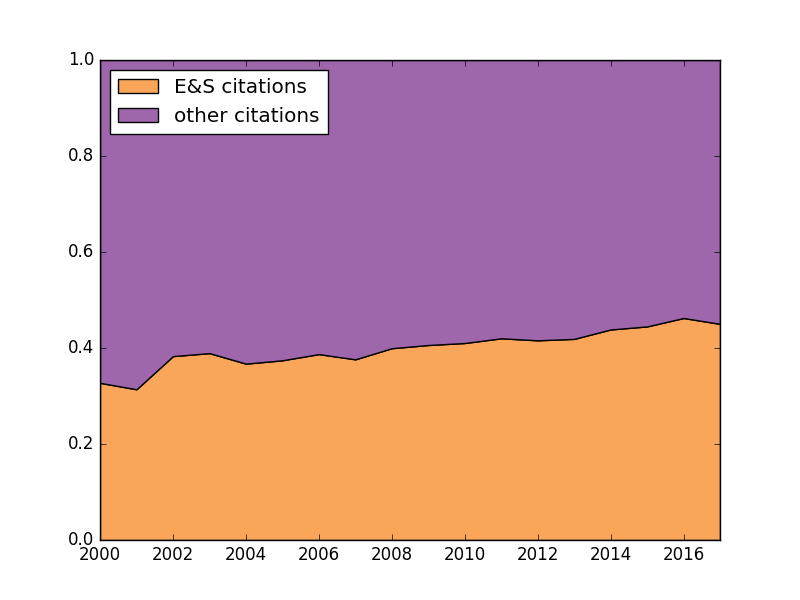}
\end{center}
\caption{
Left: share of articles published by E\&S and by other editors. Right: share of citations received by articles published by E\&S and by other editors. 
}
\label{f.article_ratio}
\end{figure}

\smallskip

The left frame of Figure~\ref{f.article_ratio} displays the evolution of the share of articles published by Elsevier and Springer (E\&S). The share of articles published by E\&S actually \emph{increases}, from about $32\%$ in 2000 to about $46\%$ in 2017. The right frame of Figure~\ref{f.article_ratio} displays the share of citations received by papers published by E\&S, which is essentially identical. Adding Wiley and Taylor and Francis to E\&S hardly changes these figures: the share of articles published by these four editorial houses goes from about $36\%$ in 2000 to about $49\%$ in 2017.


\begin{figure}
\begin{center}
\includegraphics[scale=0.33]{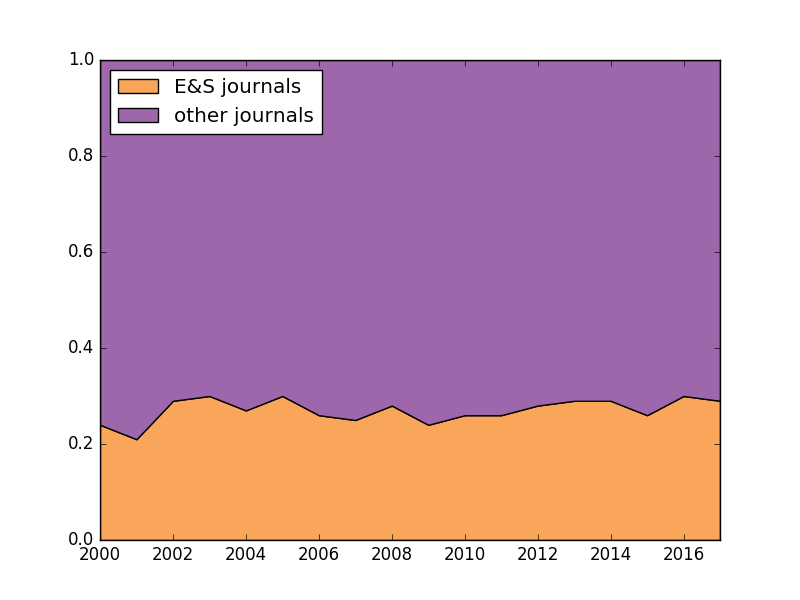}  \hspace{.1cm}
\includegraphics[scale=0.33]{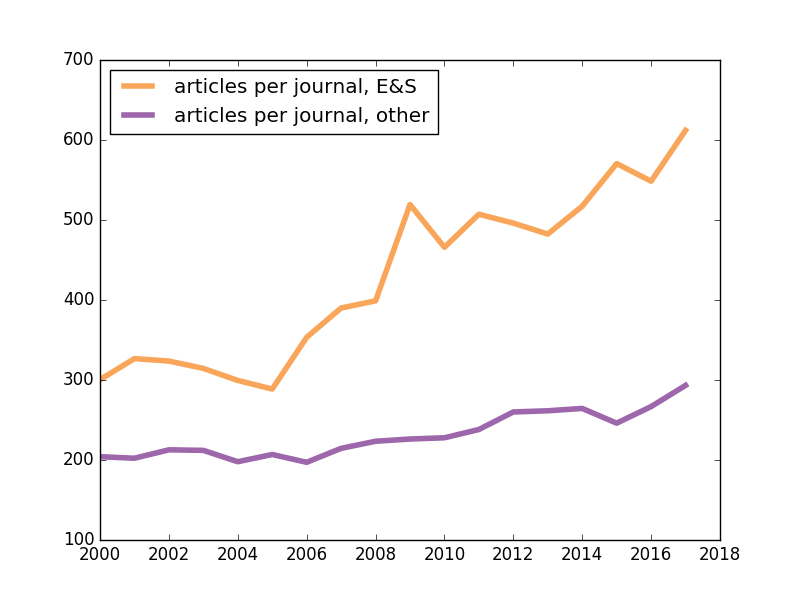}
\end{center}
\caption{
Left: share of journals published by E\&S and by other editors. Right: number of articles per journal, averaged over all journals published by E\&S, or over all journals published by other editors.
}
\label{f.journal_volume}
\end{figure}

\smallskip

Looking instead at the share of \emph{journals} (instead of articles) published by E\&S, which is displayed on the left frame of Figure~\ref{f.journal_volume},  reveals a different picture. Indeed, the share of journals published by E\&S fluctuates in the range of $20-30\%$, with no clear trend in a particular direction. As can be seen most clearly on the right frame of Figure~\ref{f.journal_volume}, journals published by E\&S have increased their volumes much more rapidly than other journals. On average, a journal operated by E\&S published about 300 articles over the years 1995 to 1999, while a journal not operated by E\&S published about 200 papers over the same period. Over the years 2012-2016, a journal operated by E\&S published more than 600 papers on average, more than doubling from 17 years back. In contrast, a journal not operated by E\&S published less than 300 papers on average, less than a $50\%$ increase from 17 years back.

\smallskip

In short, journals operated by for-profit editors have increased their volumes much more rapidly than other journals. A listing of the journals with the largest volumes confirms this analysis: over the period 2012-2016, the six journals that published the most papers are: \emph{Journal of computational physics} (3014 articles), \emph{Journal of differential equations} (1930 articles), \emph{Advances in mathematics} (1654 articles), \emph{Computer methods in applied mechanics and engineering} (1499 articles), \emph{Journal of functional analysis} (1397 articles), and \emph{Communications in mathematical physics} (1342 articles). The first five journals are edited by Elsevier, the last one by Springer.

\smallskip

As a final note, I would like to recall that the data I used is not the most up-to-date one could imagine to have, and involves an averaging over several years past. Indeed, the last data point on the graphs is labeled as 2017, and this data point in fact displays a number (of articles, or citations, etc.) that corresponds to an average over the period 2012-2016. A particularly high point of the criticism towards the practices of for-profit editors was the petition \cite{cok} initiated in 2012. Unfortunately, it is not possible to draw reliable conclusions concerning the impact of any activity that happened in the period after 2012.





\small
\bibliographystyle{abbrv}
\bibliography{ES_share}

\end{document}